\documentclass[10pt,conference]{IEEEtran}

\def\IEEEkeywordsname{Keywords}

\usepackage[font=small,belowskip=-1.1em]{caption}
\DeclareCaptionLabelSeparator{iaria}{.\quad}
\def\figurename{Figure}

\usepackage{dblfloatfix}

\usepackage{cite,breakurl}

\usepackage{amsmath,amssymb,amsfonts}
\usepackage{graphicx}

\usepackage{paralist}

\usepackage[hyphens]{url}
\usepackage{hyperref}

\title{\textbf{\Large Assessing Privacy Policies with AI:\\[-0.5em] Ethical, Legal, and Technical Challenges}}

\author{
\IEEEauthorblockN{
Irem Aydin\IEEEauthorrefmark{1}\IEEEauthorrefmark{3}, 
Hermann Diebel-Fischer\IEEEauthorrefmark{1}\IEEEauthorrefmark{3}, 
Vincent Freiberger\IEEEauthorrefmark{2}\IEEEauthorrefmark{3}, \\
Julia Möller-Klapperich\IEEEauthorrefmark{1}\IEEEauthorrefmark{3}, 
Erik Buchmann\IEEEauthorrefmark{2}\IEEEauthorrefmark{3}, 
Michael Färber\IEEEauthorrefmark{4}\IEEEauthorrefmark{3}, 
Anne Lauber-Rönsberg\IEEEauthorrefmark{1}\IEEEauthorrefmark{3}, 
Birte Platow\IEEEauthorrefmark{1}\IEEEauthorrefmark{3}}
\and
\IEEEauthorblockA{\IEEEauthorrefmark{1}Faculty of Arts, Humanities and Social Sciences, TU Dresden, Germany, Email: \{firstname.lastname\}@tu-dresden.de,\\
\IEEEauthorrefmark{4}Faculty of Computer Science, TU Dresden, Germany, Email: \{firstname.lastname\}@tu-dresden.de\\
\IEEEauthorrefmark{2}Leipzig University, Germany, Email: \{firstname.lastname\}@uni-leipzig.de\\
\IEEEauthorrefmark{3}Center for Scalable Data Analytics and Artificial Intelligence (ScaDS.AI) Dresden/Leipzig, Germany\\
}}

\begin{document}
\maketitle

\bstctlcite{BSTcontrol}

\begin{abstract}
The growing use of Machine Learning and Artificial Intelligence (AI), particularly Large Language Models (LLMs) like OpenAI’s GPT series, leads to  disruptive changes across organizations. At the same time, there is a growing concern about how organizations handle personal data.  Thus, privacy policies are essential for transparency in data processing practices, enabling users to assess privacy risks. However, these policies are often long and complex. This might lead to user confusion and consent fatigue, where users accept data practices against their interests, and abusive or unfair practices might go unnoticed. LLMss can be used to assess privacy policies for users automatically.
In this interdisciplinary work, we explore the challenges of this approach in three pillars, namely technical feasibility, ethical implications, and legal compatibility of using LLMs to assess privacy policies. Our findings aim to identify potential for future research, and to foster a discussion on the use of LLM technologies for enabling users to fulfil their important role as decision-makers in a constantly developing AI-driven digital economy.
\end{abstract}


\section{Introduction}
\label{sec:intro}
Currently, we observe a strong increase~\cite{reuters2023} in use cases and organizations using machine learning and artificial intelligence (AI), particularly Large Language Models (LLMs) like OpenAI’s GPT series~\cite{OpenAI2023}. This has many advantages for organizations~\cite{chui2023economic}. However, it is likely to increase the number and complexity of different services each user interacts with on a daily basis. It also makes it increasingly important to evaluate social, ethical, and fairness-related aspects of using personal information as input for AI-driven business processes. 

While the EU's new AI Regulation~\cite{aiact} covers numerous risks associated with the use of AI, the provisions of the General Data Protection Regulation (GDPR)~\cite{eu2016regulation} adopted in 2016 remain the legal basis for the protection of personal data. Therefore, privacy policies are an important resource. \textit{Privacy policies} describe the privacy notices that companies are making public on their websites to communicate their data processing activities on personal data. \textit{Processing} covers any (set of) operations on personal data, such as a collection, recording, organisation, structuring, storage, adaptation or alteration. Privacy policies are intended to balance information asymmetries between companies as \textit{data controllers} and individual users as \textit{data subjects} by providing them with information to assess the privacy risks and enable them to make informed, autonomous decisions. 

However, privacy policies are also known to be complex~\cite{becher2021law}\cite{warzelgoogle}, and tend to grow over time~\cite{shvartzshnaider2018analyzing}. This leads to a lack of understanding from users, and the inability to decide whether or not they prefer to consent to a certain use of their personal data; hence \textit{notification fatigue}, \textit{consent fatigue} or \textit{consent desensitisation}~\cite{Schermerconsentfatigue}. 
For example, 25\% of all Americans are asked to agree to at least one privacy policy daily, 67\% of them report to understand little to nothing about what companies are doing with their personal data, and 56\% skip reading privacy policies altogether~\cite{mcclain23americans}. In Europe, 47\% of 2,600 study participants~\cite{cisco19consumerprivacy} expressed notification fatigue due too many privacy-related notifications. It is understandingly easier for users to accept all data processing activities, even if it is against their interests. In consequence, potentially abusive or unfair data processing practices may remain unnoticed.

While the providers of services meet the formal criteria with their privacy policy, no satisfactory situation can be established for the user if no informed decision is actually made via complex data protection declarations. In this respect, empowerment, or the creation of user autonomy, respectively, remains an open task. Achieving user autonomy and `true' informed consent poses the challenge that ultimately the user has to gain knowledge of what the privacy policy entails. %

An automated assessment of privacy policies can mitigate the situation. Earlier approaches for evaluating privacy policies using machine-readable privacy policies~\cite{p3p}, classical machine learning or AI~\cite{del2022systematic,sanchez2021automatic,amaral2021ai,torre2020ai,hamid2023genaipabench} exist. However, those approaches had a focus on formal aspects such as comparing policies with given preferences, identifying deletion periods or ensuring completeness of mandatory information. We believe that LLMs can be a game changer in assessing privacy policies according to the users privacy wishes and needs. 

For example, an LLM like GPT-4o~\cite{openai2023gpt4} could be fed with a system prompt as shown in Figure~\ref{fig:prompt}, and a user prompt containing a privacy policy. It would then provide an assessment of the privacy policy along with an explanation and a rating on a five-point scale, which helps a user to identify important content, missing information and risky practices within the policy and thereby foster an informed decision.

\begin{figure}[htb]
    \centering
\fbox{\parbox{.9\linewidth}{\small %
\em
You are a critically scrutinizing, experienced authority for data privacy, and an expert on assessing privacy policies. You have 20 years of experience in consumer protection, data protection, cybersecurity and related fairness aspects. You have already provided precise expert opinions in many court proceedings. Your task is to uncover aspects in privacy policies that are ethically,
morally or legally questionable. Shortly explain your thoughts and how you came to your conclusion. In particular, point out potential risks to users. IMPORTANT: Also rate the privacy policy for each of your criteria on a 5-point Likert scale. Check whether the following privacy policy is fair and ethical towards its users. The user wants a short and concise critical review and an assessment of vulnerabilities in privacy policies. IMPORTANT: You must answer in less than 300 words.}}
    \caption{System prompt to assess a privacy policy}  
    \label{fig:prompt}
\end{figure}

\newcommand{\myspc}[0]{\\[-1.75mm]}

\begin{figure}[htb]
    \centering
\fbox{\parbox{.9\linewidth}{\small %
\em
\textbf{\normalsize Amazon.com Privacy Notice Review}\myspc

\textbf{1. Transparency and Understandability:} The policy is clear, well-structured, and provides links for more detailed information, making it user-friendly. However, users may find the comprehensive format overwhelming. {\footnotesize \em (Rating: 4/5)}\\
\textbf{2. Data Collection Scope:} Amazon collects extensive data, from basic contact information to behavioral data via cookies. This vast data collection is concerning, as it may lead to invasive profiling and privacy risks. {\footnotesize \em (Rating: 2/5)}\\
\textbf{3. Data Usage:} While Amazon provides clear reasons for data usage, including service improvement and personalized ads, the broad scope can be intrusive. It should ensure by providing clear opt-out methods for users. {\footnotesize \em (Rating: 3/5)}\\
\textbf{4. Third-party Sharing:} Data sharing with third parties, though typical, poses risks especially concerning how those parties handle data. The policy does ensure some protective measures but remains ambiguous about compliance monitoring. {\footnotesize \em (Rating: 3/5)}\\
\textbf{5. Security Measures:} Amazon mentions robust encryption and compliance with PCI DSS, which reflects strong security practices. However, the responsibility placed on the user for password protection is emphasized. {\footnotesize \em (Rating: 4/5)}\\
\textbf{6. User Autonomy (Consent \& Control):} Amazon provides mechanisms for users to access, rectify, and delete their data. The range of consents, from opting out of ads to withdrawing permissions, shows a commitment to user choice, albeit complex. {\footnotesize \em (Rating: 4/5)}\\
\textbf{7. Children’s Privacy:} Good measures in handling children's data, restricting services to adults, and requiring parental consent. {\footnotesize \em (Rating: 4/5)}\myspc

\textbf{Potential Risks to Users:}\\
{\em (a) Data Profiling:} Extensive data collection can lead to detailed profiling, risking user autonomy and privacy.\\
{\em (b) Third-party Data Handling:} Despite contractual protections, third parties’ data mishandling poses %
risks.\\
{\em (c) Complex Opt-out:} The user control mechanisms, while present, may be too cumbersome.\myspc %

\textbf{Final Assessment:} While Amazon’s privacy policy demonstrates strong commitment towards transparency and user rights, the vast scope of data collection, potential third-party risks, and complexity in user control indicate areas needing improvement. {\footnotesize \em (Overall Rating: 3.5/5)}}}
    \caption{GPT-4o’s assessment of Amazon's privacy policy}  
    \label{fig:result}
\end{figure}

Figure~\ref{fig:result} shows an exemplary analysis of the privacy policy of Amazon.com, using the GPT-4o model and the prompt from Figure~\ref{fig:prompt}. This confirms that such an assessment is technically feasible. However, it leaves numerous open questions:
From a technical perspective, it is difficult to determine the correctness of the LLM assessment. From a user-centric point of view, users with different educations might perceive the assessment of the LLM differently, and could also decide to blindly trust/distrust %
an AI assessment. Because the LLM/prompt developer has a large influence on how empathetic/objective the assessment is generated by the LLM, there are open moral questions. Ethical questions arise, when a user loses the ability to understand a privacy policy because an LLM offers a simpler solution. From a legal perspective, it can be questioned whether a user's consent based only on the information provided by the LLM, meets the requirements of Art. 7 GDPR, and how to handle situations where a malfunctioning or misuse of the LLM causes actual harm. Additionally, a privacy-enhancing AI tool must comply with traditional legal norms, and the new AI~Act~\cite{aiact}.

In this paper, we explore the challenges of assessing privacy policies with an LLM from an interdisciplinary point of view. %
In particular, we consider three pillars: 
\begin{compactitem}
\item \textit{Ethical:} We explore the ethical, moral, and fairness-related challenges that arise when the assessment of privacy policies is no longer in the hands of the users, but in the hands of a very convincing LLM that could be wrong or biased, or using biased training data. 
\item \textit{Legal:} We discuss the extent, to which the use of an LLM for a privacy policy assessment is or can be made compatible with existing legal norms. We also consider how such norms need to change to enable AI-supported decision-making and thereby ensure consumer welfare and self-determination.
\item \textit{Technical:} 
We investigate how to assess privacy policies automatically by using LLMs, and we examine how to use natural language processing, machine learning, linguistics, knowledge representation, etc. to evaluate the correctness, understandability and applicability of the LLM assessment. 
\end{compactitem}

To the best of our knowledge, we are the first to explore with an interdisciplinary approach, 
how LLMs can be used to assess %
the practices documented in privacy policies. For this research, we combined the expertise of our groups 
\textit{Civil Law, Intellectual Property, Media and Data Protection Law;
Data Privacy and Security;
Practical Theology with a focus on Religious Education, Ethics;
Scalable Software Architectures for Data Analytics and
Systematic Theology} 
from the universities of Dresden and Leipzig. 
This was a challenge, because each of these groups has its own research culture, which had to be combined in order to achieve interdisciplinary results.
Our objectives are particularly relevant, because politics and jurisprudence are still in the process of identifying options to ensure user privacy in the AI era. 

\textit{Paper Structure:} The next section reviews related work. Section~\ref{sec:method} describes how we systematically derive challenges for assessing privacy policies with LLMs. In the Sections~\ref{sec:chal-tech} -- \ref{sec:chal-legal}, we will compile our set of challenges. Section~\ref{sec:discussion} contains a discussion. The paper concludes in Section~\ref{sec:conclusion}.
\section{Related Work}
\label{sec:related}

This section contains a review of literature, which we derive our challenges from in the following sections. 

\subsection{Legal Background}
\label{subsec:priv}

The GDPR~\cite{eu2016regulation} aims to provide data subjects in the European Union with control over data processing activities, which could potentially impair their fundamental rights. The regulation aims to enable processing where it is necessary and in line with the objectives of the data subject and to prevent it unless the risks associated with the processing are outweighed by a corresponding benefit in the public interest or in the interest of the data subject~\cite{malgieri2020concept}. These goals are reflected in the requirements for lawful processing in Art. 6 GDPR as well as in the requirements for effective consent in Art. 7 GDPR. 

While the EU's efforts to establish a digital single market, like the AI~Act~\cite{aiact} and the Digital Services Act (DSA)~\cite{euDSA}, aim to set high standards by prohibiting certain harmful practices involving the use of personal data, they still leave a great degree of flexibility for user autonomy to consent.  
Such consent is possible when the data subject is capable of making an informed decision~\cite{trustworthyaiguideline} autonomously. This requires sufficient knowledge to evaluate advantages and disadvantages. 
Therefore, knowledge should be presented to the person understandably and transparently~\cite{Zaeem2020}, hence the need for privacy policies. The GDPR~\cite{eu2016regulation} requires data controllers to make their privacy policies complete, readable, and possible to understand for all kinds of `typical users' of a service, which can be persons with different abilities or knowledge. 

AI-based applications and services integrating them, add another layer to data processing practices, which users need to comprehend. 
Such applications might force users to disclose more personal data and impose difficulties in assessing the costs and benefits from the user's viewpoint due to a lack of transparency or understandability. 
In 2008, it was calculated that it would take an average internet user between 181 and 304 hours every year to read every privacy policy of all web services they are using~\cite{mcdonald2008cost}. The GDPR~\cite{eu2016regulation} also caused the complexity and length of privacy policies to be increase~\cite{shvartzshnaider2018analyzing}. Furthermore, the extensive use of personal data to train AI and its unpredictable outcome increases the potential of significant impairment of the user's needs and interests~\cite{JMK_Sicherheit}.

Different approaches to support the users handling their personal data have been discussed in the legal literature~\cite{Sattler2022}. It is important to highlight that achieving transparency in privacy policies is an ongoing collective effort, and simplifying tools to achieve this purpose have been experimented on with privacy icons, a machine-readable label system~\cite{Cranor2012NecessaryBN}, one-page summary of the privacy policy, privacy taxonomy or a `privacy nutrition label' as well as different kinds of technically supported privacy management systems (PIMS) or the deployment of data trustees~\cite{Ulmenstein}. However, there are also findings that even the simplified declarations of data processing practices may not change user behaviour to disclose intrusive information~\cite{simplificationtest}.

\subsection{Ethics}
\label{subsec:ethics}
The definition and continuous discussion of privacy ethics~\cite{mainz2021indirect}\cite{lundgren2020dilemma}\cite{elliott2022ai}\cite{Marmor2015} demonstrates the need for an evaluation to be comprehensive, coherent, systematic, and logical in its reasoning.
Codes of ethics, legal statutes, or international declarations embody norms and values ingrained in our society. They can provide helpful input for an ethics assessment of new and emerging technologies~\cite{harris2011ethical}. However, the field of ethics misses strict definitions and straightforward tools. Owing to that, ethics must not be understood as a tool to solve moral problems but should be regarded as a way to describe, understand, and reflect on them.
One common issue is transparency in privacy policies~\cite{transparency}\cite{belcheva2023understanding}\cite{becher2021law}\cite{reidenberg2015disagreeable}. Policies tend to be long, written in inaccessible language, and users tend to struggle to understand their content, 
resulting in issues like consent fatigue~\cite{choi2018role}. Privacy policies can also use persuasive language~\cite{belcheva2023understanding}\cite{pollach2005typology} to let the users trust a service that, for instance, claims all rights over users’ data~\cite{koetsier2019viral}.

The given issues have motivated privacy assistants~\cite{tesfay2018privacyguide}\cite{nokhbeh2020privacycheck}\cite{amaral2021ai}. 
Emerging capabilities by scaling up LLMs~\cite{wei2022emergent} have given them a wide range of applicability~\cite{guo2023close}. This makes them interesting as a tool for assessing privacy policies~\cite{hamid2023genaipabench}\cite{tang2023policygpt}\cite{palka2023no}. 
Seeking privacy serves two fundamental purposes: security interests (stay unharmed) and privacy per se~\cite{elliott2022ai}. Privacy per se is about managing how we show ourselves to the outside world and, more broadly, about our autonomy~\cite{elliott2022ai,Marmor2015}. Privacy and the right to privacy are two different concepts in the philosophical debate on privacy ethics. Privacy may be infringed upon, but not the right to privacy, depending on the circumstances surrounding the collection of personal data about oneself by a third party and the underlying intention~\cite{Marmor2015}.
Privacy ethics addresses access others have to one’s information as well as control one has over it~\cite{decew1986scope}. Complex privacy trade-offs and the balance of power between the data controller and data subject are relevant to the discussion~\cite{acquisti2015privacy}\cite{Wilsdon2022}.
While concepts such as `fairness' are deemed important, it remains difficult to define them precisely. As they are lacking a precise definition, the operationalization of these concepts, i.e. transferring them into a model and making them computable—remains an obstacle~\cite{geldhauser2024diverse}\cite{diebel2023technisch}\cite{huppenbauer2021}. To overcome this problem, the unavoidable operationalizations need to be lined out and explained, and ideally be user-configurable. 
 
Furthermore, concerns about surveillance~\cite{elliott2022ai}\cite{benn2022s}, choice impact~\cite{choi2019privacy}, manipulation~\cite{Marmor2015}\cite{Wilsdon2022}, and power imbalances~\cite{Wilsdon2022}\cite{acquisti2015privacy} have been brought up in the context of online user privacy.
To address these issues, user education on such problems is required~\cite{platow2021a}\cite{platow2021}. The user needs to be aware of the underlying ethics-related problems and the way these have been `solved' in terms of implementation into such a system.

Regarding an ethical assessment of privacy policies, it is required that ethics assessments  consider all perspectives with their normative grounding~\cite{gogoll2021ethics}. Ethical aspects may come into effect unintendedly, sometimes as second-order consequences~\cite{gogoll2021ethics}. Thus, the consistency of a moral assessment provided by the model is mandatory as it influences user judgment~\cite{krugel2023chatgpt}. 
A good assessment should be concise and understandable. Thus, explainable AI employing various metrics should be considered~\cite{nauta2023}.  Informational fairness~\cite{freiberger2024legal} should be considered as well as addressing privacy trade-offs and power imbalances is important~\cite{acquisti2015privacy}\cite{Wilsdon2022}.
In addition, moral psychology can be used to study LLMs~\cite{hagendorff2023machine}. This entails investigating potential biases in the model's representation of moral judgments and moral reasoning, as well as to what degree they are present in the model's outputs. 
For example, it has been discovered that ChatGPT's moral guidance is inconsistent when presented with a moral dilemma~\cite{krugel2023chatgpt}.

\subsection{Large Language Models and Prompting}
\label{subsec:prompting}
LLMs iteratively predict the next token to produce text for a given query. The GPT-4 models are a series of capable LLMs introduced by OpenAI in 2023~\cite{achiam2023gpt}. GPT-4 already has been shown to have a set of reasoning capabilities~\cite{espejel2023gpt}\cite{bubeck2023sparks}.

LLMs can be applied in many different domains~\cite{bubeck2023sparks}\cite{guo2023close}. To improve model performance and mitigate limitations plenty of prompting strategies have been developed~\cite{rae2021scaling}\cite{kojima2022large}\cite{wei2022chain}\cite{brown2020language}\cite{yao2024tree}\cite{zhou2022large}. For a review on basics of prompting LLMs, we refer to~\cite{liu2023pre}.

Reasoning-related tasks benefit from prompt-engineering strategies~\cite{rae2021scaling}\cite{kojima2022large}. Such strategies are referred to as Chain-of-Thought prompting~\cite{wei2022chain} (asking the LLM step by step), Reflection~\cite{kim2024language} (asking the LLM to rethink its answer), Few-Shot prompting~\cite{brown2020language} (giving examples) or Repetition~\cite{zamfirescu2023johnny} (repeating relevant aspects in the prompt). Over-generalization is a common issue in prompting LLMs~\cite{zamfirescu2023johnny}.
Controlled small prompt modifications can largely affect the model's output~\cite{hagendorff2023machine}. 
The seed and other model parameters must be fixed if repeatable results are important~\cite{lee2023}.
By rephrasing the prompt, robustness can be evaluated~\cite{hamid2023genaipabench}\cite{jiang2020can}\cite{yuan2021bartscore}.

\subsection{Automatic Text Analysis and Assessment}
\label{subsec:tools}

Assessing privacy policies is part of automatic text assessment based on natural language processing (NLP) techniques. 
Automatically assessing text using readability metrics started with text statistics (e.g., word frequency, word length, sentence length). Readability metrics are language-specific~\cite{10.1162/00398}. The most popular~\cite{article-readability} metrics are the  \textit{Gunning fog index}~\cite{gfi}, the \textit{Flesch reading ease}~\cite{kincaid1975derivation}, and the \textit{Simple Measure of Gobbledygook}~\cite{mc1969smog}. The metrics have been criticized for their inability to capture more complex aspects of a language. 

Assessing the readability with machine learning~\cite{crossley2017predicting}\cite{10.1162/00398} generally produces better results than the traditional approaches, but requires more effort, such as creating data sets and training the model. Specifically, for the German language, methods that use traditional language models~\cite{hancke-etal-2012-readability}\cite{blaneck-etal-2022-automatic} or which use semantic networks in comparison to simple surface-level indicators to calculate text readability~\cite{inproceedings} were created. Despite~\cite{DBLP:conf/ecir/PickelmannFJ23}, to our knowledge, there is no work focusing on building a system for the automatic assessment of text readability in German~\cite{DBLP:conf/ecir/PickelmannFJ23} use the pre-trained language model BERT for assessing the readability of text.

Texts can also be analyzed for sentiment~\cite{guhr-EtAl:2020:LREC}. %
In the context of media bias detection, texts have been automatically assessed with respect to different aspects of bias~\cite{newsbias-cikm-2020}: (1) Hidden Assumptions and Premises, (2) Subjectivity, (3) Framing and (4) Overall Bias.

\subsection{Automated Assessment of Privacy Policies}
With the commercialization of the Internet, the number of privacy policies that had to be read increased. Thus, there is a long history of attempts to assess privacy policies automatically. 
To name a prominent example, in 2004, the Platform for Privacy Preferences (P3P)~\cite{p3p} standardized a protocol, that allowed data controllers to publish machine-readable privacy policies. Web browser plug-ins such as the Privacy Bird~\cite{cranor2003p3p} allowed users to specify their privacy preferences, which were automatically compared against P3P policies. However, websites were not obligated to use P3P, and the specification of meaningful preferences is a difficult task. Later approaches used NLP approaches such as morphological, lexical, syntactic, and semantic analyses or ontology reasoning to assess privacy policies (see~\cite{del2022systematic} for a detailed comparison). For example~\cite{sanchez2021automatic}, a  support vector machine can be trained to map the sentences of a privacy policy to the mandatory information and user rights, that must be declared in a privacy policy. AI has also been used, e.g., to verify whether the content of a privacy policy is complete~\cite{amaral2021ai}\cite{torre2020ai} according to the GDPR. 
The advantage of such approaches is that they work without the help of data controllers. The work closest to ours is~\cite{hamid2023genaipabench}, which gauges the effectiveness of ChatGPT-4, Bard, and Bing AI for assessing privacy policies. This approach acknowledges the technical feasibility of such an assessment and provides quality measures, but leaves aside the impact of the prompt engineer and all ethical and legal issues.

\section{Our Research Approach}
\label{sec:method}

To provide a systematic overview of ethical, legal, and technical challenges of letting an LLM assess privacy policies for the users, we pursued an explorative approach. This approach is the common basis of the respective research methods from our very different research disciplines, which enables us to combine our findings into an interdisciplinary result: 

\begin{compactenum}
    \item As a first step, we \textbf{compile an annotated bibliography} on LLMs and privacy policy assessment (cf. Sec.~\ref{sec:related}). 
    
    We also \textbf{implemented and tested a number of approaches} for assessing domain-specific texts with an LLM to gain first-hand experience, and discussed these in our research groups, e.g., \cite{freiberger2024legal}\cite{juettner2023chatids}\cite{freiberger2024balancing}\cite{hoffmann2024ChatSEC}. 
    \item In a second step, we use this combination of background information and first-hand experience to \textbf{formulate a series of challenges} in the three pillars described. 
    \item Finally, we \textbf{filter for challenges} that are specific to our application domain, i.e., we exclude general difficulties in obtaining training data, performance, explainability, enforcement of legal norms for complex IT systems, etc. 
\end{compactenum}
In the following sections, we describe the challenges we have obtained using this approach.

\section{Technical Challenges}
\label{sec:chal-tech}
Our research groups have first-hand experience in investigating the transparency of German privacy policies~\cite{transparency}. We also investigated, to which extent LLMs can be used to assess and explain difficult security issues~\cite{juettner2023chatids} or solve exam questions at Bachelor's or Master's level~\cite{buchmann2023chatgptexams}. We also modeled privacy practices as structured design patterns~\cite{robak2019model}\cite{robak2019extract}, the implementation quality of which can be estimated or measured. With a fairness certification for NLP and LLMs, we introduced criteria for addressing biases in the model output~\cite{freiberger2024faircert}. We also let an LLM assess multiple dimensions of fairness in privacy policies~\cite{freiberger2024legal}\cite{freiberger2024balancing}. Based on this hands-on experience and our literature base in Section~\ref{sec:related}, we derive six technical challenges:

\textbf{Interest in certain privacy policies is private data.}
When using an LLM to assess a privacy policy according to ethics, morale, legality etc., the users reveal their interests. Assume a user calls for an assessment of a privacy policy of an AI company, and expresses concerns due to personal data used as training data. This is sensitive information, which %
calls for anonymization or on-premise solutions. 

\textbf{Annotated data sets as a ground truth are limited.}
To enable few-shot prompting and fine-tuning of LLMs, as well as to enable a solid evaluation, annotated data sets with ground truths for policies are required. The closest to such a data set is TOSDR~\cite{tosdr}, which provides crowd-sourced / automatically generated annotations for popular policies. However, these annotations are limited in number and quality control.

\textbf{The assessment requires individual prompts.}
LLMs need to incorporate user-specific preferences and concerns into their prompts to effectively assess privacy policies, accommodating various social and educational backgrounds. This includes fighting biased responses, ensuring the representation of diverse opinions. An interactive tool or a set of tailored prompt templates that can handle these nuances may be required. %

\textbf{Explaining the assessment to the user.}
Due to hallucinations of the LLM, the LLM misunderstanding a privacy policy, or a user misunderstanding the LLM assessment, mistakes might occur. It is an important challenge to structure an LLM approach for the assessment of privacy policies in a way that tolerates mistakes. For example, an interactive approach might allow the LLM to ask back for specific user preferences, which increases the user's awareness.

\textbf{The data controller must not influence the assessment.}
Companies might utilize limitations regarding the robustness of an LLM-based ethics assessment to get more favorable assessments without improving their privacy practices. Targeted variations of their policies without changing their semantic content could be used to optimize for higher ratings. This makes testing robustness of the LLM output, and particularly adversarial testing of LLM assessments essential.

\textbf{It needs strategies for consistent LLM assessments.} 
The LLM assessment may depend on nuances in the prompt and the privacy policy, that a human would overlook, resulting in different assessments for similar privacy policies. This undermines the reliability and trustworthiness of generated assessments. Therefore, approaches for enforcing structure in outputs and optimizing the prompting in a way that is consistently followed by the LLM are important.

\section{Ethical Challenges}
\label{sec:chal-eth}

Incorporating ethical analysis into an AI environment requires an operationalization of ethics with the aim of its mathematical implementation. We delved into concepts such as fairness in AI from a quantitative perspective~\cite{geldhauser2024diverse}. We also analyzed the relationship between technology and ethics. In the era of AI, this relationship requires human oversight to avoid a blind and potentially misleading technization of otherwise qualitatively expressed goals~\cite{diebel2023technisch}. With this in mind, we formulate four challenges: 

\textbf{Different stakeholders have different objectives.}
It is an open issue how to assess the dimensions of fairness for privacy policies: While the users should want to provide as little data as possible, the data controllers rely on data to further develop their services. It is challenging to identify a way to balance these objectives responsibly.

\textbf{Identifying socially desirable practices.}
While legal regulations hint at what voters might want as they are results of law-making processes in representative democracies, it remains opaque, what is socially desirable. 
Thus, what would a set of data management practices look like that meet everyone's privacy needs, and can serve as a reference for assessing privacy policies?

\textbf{Operationalizing the evaluative criteria.}
To identify a socially desirable outcome, methods are needed to operationalize evaluative criteria for the assessment of privacy policies. 
While the same setting might be judged as fair or unfair depending on the position of those who judge, a metric can objectively measure intersubjectively acceptable parameters. However, such parameters are unknown yet.

\textbf{The status of the assessment must be defined.}
The LLM's assessment is meant to guide the user regarding the acceptability of a privacy policy. 
The output of the LLM is the product of intricate `translation processes'~\cite{diebel2023technisch}, in which the quantitative elements (`scores') are expressed in words. It can be paternalism, if users take the LLM's assessment as their decision, narrowing down the user's autonomy. However, the LLM could be the only option to quickly assess privacy.

\section{Legal Challenges}
\label{sec:chal-legal}

Our previous research in fields such as transparency, trust, data protection, and responsibility when using AI in legally sensitive areas~\cite{JMK_ChatGPT} has shown that, in particular, transparency and user trust are essential to ensure broad acceptance and fair use of AI technologies~\cite{lauber2022kunstliche}. In addition, property rights, data protection~\cite{lauber2019handbuch}\cite{JMK_Sicherheit} and liability~\cite{JMK_KI_Wertschöpfung} must be clearly regulated and respected in the digital age to protect the rights of individuals. On this basis, we identified four challenges: 

\textbf{Common understanding of fairness and transparency.} 
While Art. 12 of the GDPR requires privacy policies to be comprehensible and comprehensive to allow autonomous decisions, 
there is a lack of common understanding of fairness and transparency and how to achieve them. Navigating the AI era requires a clear understanding from all stakeholders~\cite{edpbguideline,edpsopinion,euaniprivacybydesign}. New approaches are needed to prove lawful processing, i.e., to describe in clear language how the data is collected, pre-processed, used for training, which prompts and outputs are used, and whether training is carried out with outputs~\cite{reportchatgpttaskforce}, mainly focusing on explainability~\cite{trustworthyaiguideline}.

\textbf{Questionable validity of AI supported consent.} Art. 7 GDPR requires the subject of data processing to make an informed and free decision. Using an AI tool as a support can, under certain circumstances, be seen as an influence jeopardising the validity of given consent. It becomes more questionable in case the AI tool is manipulative, which is prohibited by Art. 5 of the AI~Act~\cite{aiact}.

\textbf{Compliance with the AI~Act and related regulations.}
An AI tool to assess privacy policies can be considered a High-Risk AI System in the meaning of Art. 6 ff. AI~Act~\cite{aiact} and therefore the provider, distributor, or deployer can be subject to extensive obligations. Depending on the specific use in a single case, it might also fall under the scope of the ePrivacy Directive~\cite{eprivacy} and its potentially varying national implementation acts, e.g., the German TDDDG~\cite{TDDDG}.

\textbf{Liability of the assessment is an open question.}
If users base their consent to the data practices of a data controller on the assessment by an AI tool, it is not clear who shall be liable in case of possible damages occurring on either side, e.g, the loss of control over personal data of the data subject or business losses of the data controller.

\section{Discussion}
\label{sec:discussion}

Using LLMs is a promising approach to analyzing privacy policies. LLMs can efficiently process numerous, lengthy privacy policies without getting tired, or losing focus and interest. This is a very practical feature, as every active Internet or smartphone user uses many different services every day, each of which has its own privacy policy. Furthermore, it might be possible to carefully instruct the LLM to produce assessments that are more consistent and objective than a human assessment. With suitable prompting techniques, it is also possible to individualize the assessment for different priorities of the users at a given time. As an example, it is possible to prompt the LLM for an assessment from a non-native person's point of view, which might include less complex words, and terminology. As the technology evolves a use by supervisory authorities to monitor compliance with the requirements of Art. 12 and 13 GDPR is conceivable.

However, existing LLMs are notorious for hallucinations. Reluctant service providers, who suspect that their customers are using LLM, might be tempted to write its privacy policy in a way that provokes such hallucinations, e.g., by using unusual phrases that were underrepresented in the training data of the LLM. In addition to these general limitations, we have identified a large number of different interdisciplinary challenges for an LLM-based assessment of privacy policies. Nevertheless, we think that it would be better to have an LLM reading privacy policies than a human who is too busy to read them at all as a preliminary step, and prevent potential privacy risks go unnoticed for sure. As human agency and oversight are key components of a trustworthy AI, it should be kept in mind that the LLM-based assessment of privacy policies aims to support the autonomous decision-making of the users to strengthen their fundamental right to privacy, not to replace the decision-making authority~\cite{trustworthyaiguideline}.

\section{Conclusion}
\label{sec:conclusion}

Effective privacy policies are essential for maintaining transparency in data usage and enabling users to assess privacy risks. However, the complexity and length of these policies can often lead to confusion and consent fatigue, where users might inadvertently agree to practices that are not in their best interest. %
To tackle these challenges, our study investigated the use of LLMs to automatically evaluate and simplify privacy policies. We explored the technical feasibility, ethical implications, and legal compatibility of using LLMs for this purpose. We aim to identify potential areas for further research and to stimulate a dialogue on how risk-based policies could be effectively shaped using LLM technology.

Employing large-scale language models to interpret and simplify privacy policies is a crucial and timely research endeavor. This interdisciplinary approach addressed the urgent need for transparency in AI-driven contexts and utilized the unique capabilities of LLMs to enhance user understanding and decision-making. By integrating legal expertise, advanced machine learning technologies, and considerations of ethical and societal impacts, our research aims to reduce consent fatigue, counter unfair data practices, and empower individuals in a digital age dominated by complex data interactions.

\bibliographystyle{IEEEtran}
\bibliography{literature}

\end{document}